# Distributed Fault-Tolerant Avionic Systems – A Real-Time Perspective

N.C. Audsley, M. Burke
British Aerospace Dependable Computing Systems Centre,
Department of Computer Science,
University of York, York Y01 5DD, UK.
{neil,baemike}@cs.york.ac.uk

*Abstract*—This paper examines the problem of introducing advanced forms of fault-tolerance via reconfiguration into safety-critical avionic systems. This is required to enable increased availability after fault occurrence in distributed integrated avionic systems (compared to static federated systems). The approach taken is to identify a migration path from current architectures to those that incorporate re-configuration to a lesser or greater degree. Other challenges identified include change of the development process; incremental and flexible timing and safety analyses; configurable kernels applicable for safety-critical systems.

TABLE OF CONTENTS

## 1. INTRODUCTION

The aerospace industry is divided into a number of sectors (Missiles, Military Aircraft, Civil Aircraft, etc). The development processes used to produce airborne computing systems are similar, at an abstract level. These processes can be described as a complex interaction between the technical definition and the procurement activities. Ideally, the technical process proceeds the procurement process, however, these often occur simultaneously, due to the long lead times on computing equipment (ruggedised for avionic use). The authors assert that this abstract process is a major source of technical problems and lifecycle costs, since hardware dependencies are introduced at an early stage. For example, the target architecture is specified to fine detail prior to software being written. The cost penalty associated with changing these early design decisions is significant. As a result these decisions become a major constraint on the down stream lifecycle activities.

One potential solution for this problem would be to serialise system refinement and procurement. This is unlikely to ever be a viable option due to current and growing commercial pressures to reduce development time-scales - hence the abstract development model of parallel technical development and procurement is likely to remain for the foreseeable future.

Another potential solution is to migrate towards distributed computing architectures, where applications are partitioned and allocated over (potentially) several computing resources. Such architectures offer flexibility which will mitigate (but not eliminate) the worst effects of the abstract systems development process. This flexibility allows mapping to specific computing resources to be delayed and modified much more cheaply and easily. Also, modular increments of total computing resource, both through development and during service, can be performed without invalidating certification / qualification evidence. To achieve this solution, a number of enabling technologies need to be available, from sufficiently fast hardware, through operating systems (that enable hardware abstraction for the application).

An important "side-effect" of the distributed approach is enhanced fault-tolerance. Typically, avionic systems use "static" fault-tolerance, where each application, together with its dedicated hardware, is replicated. Distributed architectures enable enhanced fault-tolerance through reconfiguration – shared spare computing resources are provided in the system which are dynamically allocated to failing applications when necessary.

The focus of this paper is the migration toward distributed architectures with enhanced fault tolerance, from the typically static architectures used in current avionic systems. The remainder of this section provides further motivation and background. In subsequent sections, we consider the specific challenges of the software development process; distributed fault-tolerant architectures; software architecture and timing analysis.

We note that the British Aerospace Dependable Computing Systems Centre is allowed visibility of the processes used across a range of avionic system suppliers. Specific process information is restricted under the terms of commercial non-disclosure agreements. This breadth of visibility has allowed the authors to assert general observations and challenge the academic community to produce counter examples to the assertions.

*Background – Traditional v Distributed Avionic Systems*

Traditional aerospace computing architectures are "federated" or "black-box" in that individual applications are allocated to distinct processing resources. Such architectures, together with the software development process used to develop applications, can lead to a number of potentially costly problems, including:
- **resource under-utilisation -** since an application may only be used during part of the mission or flight, its associated resource is only partially utilised.
- **inflexible fault-tolerance -** fault-tolerance is often achieved by voting between multiple "computing lanes" which are identical in function. Typically, recovery from transient failures is possible. However, once a permanent fault condition is detected in a lane, it is shutdown. No attempt is made to reconfigure the system to bring back a failed application.
- **expensive timing analysis -** each application is structured as a cyclic schedule, with interactions between applications performed in a similar cyclic manner across some communications media. Static schedules are difficult to change and analyse with respect to their timing properties. The problem is compounded when attempting to analyse the timing properties of communicating (distributed) applications.
- **expensive maintenance -** changes to the software and/or hardware are problematic. Typically, entire systems need to be re-analysed for timing performance, re-tested and re-certified, even for a minimal change.

As suggested earlier, one potential solution for future avionic systems is the use of a distributed integrated architecture [1,2]. Here, applications are structured such that parts of an application may be allocated to different processing resources. This enables:
- **efficient use of computing resources** - achievable by enabling different applications to be split amongst different processing resources. Also, scheduling techniques, such as fixed-priority, enable more efficient resource usage.
- **fault recovery via reconfiguration -** the flexibility provided by distributed integrated architectures enables reconfiguration. Note that reconfiguration is difficult in federated systems.
- **evolutionary not revolutionary software / hardware changes -** achieved via technology transparency to enable exploitation of latest technologies. This requires kernel technologies with hardware independent interfaces. Portable code techniques can also be employed. Such technology transparency allows qualification evidence to be re-used, without completely re-generating it.

The flexibility given by integrated architectures provides:
- **increased resilience to faults -** systems are generally more available, since they reconfigure "new" copies of failed applications onto spare processors. This is increasingly important as:
  - the amount of total software in avionics systems is increasing;
  - the proportion of safety critical software is increasing;
  - the percentage of safety critical software expected to fail operational is increasing.
- **easier static analysis of applications and systems -** with respect to temporal and safety properties;
- **reduced maintenance cost –** since systems can reconfigure to exclude a faulty component, service intervals may be increased.

These may generate cost saving over the product lifecycle (typically 30 years in the aerospace sector).

*Background – Enabling Technologies*

Migration to distributed architectures has been hampered by the lack of a number of inter-related enabling technologies. Delivering these enabling technologies through a combination of in-house R&D and exploitation of advances provided by the commercial computing and telecommunications industry is now an achievable goal. Use of these more distributed integrated architectures is now a viable mitigation strategy for the problems induced by the abstract systems development process described above.

The use of distributed integrated architectures requires key enabling technologies to be available:
- **high speed processors and communications technologies -** given the need for extremely fast computation and communication – e.g. the desirable performance advantages of an aerodynamically unstable aircraft can only be realised safely where the flight control system is guaranteed to achieve a defined high frequency response.
- **suitable development process –** current development processes are optimised for the development of federated systems and do not exploit the benefits offered by distributed architectures. As noted earlier, for commercial reasons, these processes are unlikely to change radically. Therefore, one challenge is to enable distributed systems development by a series of incremental changes to the current process.
- **fault-tolerant distributed architectures –** such architectures must provide sufficient fault-resilience to reduce design and maintenance cost, to counterbalance the inevitable increased initial certification cost.
- **configurable safety-critical kernels [3] –** kernels allow software development with minimal dependence on the underlying hardware; support independence from specific compiler or code generator tools; enable integration of applications written in different languages. This may reduce long term maintenance costs, and support migration from existing aerospace products. We note that available commercial operating systems are, in general, not suitable for the exacting requirements of the aerospace sector.
- **timing analysis [4,5] and safety analysis [6] that can be applied at many stages of the life-cycle** – static analyses, including those for both timing and safety, need to be scalable and incremental to reflect the nature of the architecture. This enables evolutionary maintenance activities to occur without requiring re-analysis of an entire system. Over the long term, the ability to reason about the performance of the application software will significantly decrease the cost of system maintenance activities. Also, the potential for reasoning about degraded modes of operation is introduced.
- **time bounded techniques for fault-detection and recovery via software reconfiguration** – fault-tolerance via software reconfiguration is considered important, to enable longer periods between maintenance; together with greater availability of full system functionality. We argue that the availability of techniques for performing reconfiguration of software enables a more flexible approach toward fault-tolerance in avionic systems. Offline reconfiguration can be used to enable a system with a hardware fault to be configured for flight so that the faulty component is not used. Dynamic reconfiguration can be used to recover the full function of a system after a failure during flight. In both cases, the full function of the system is not lost due to a failure.

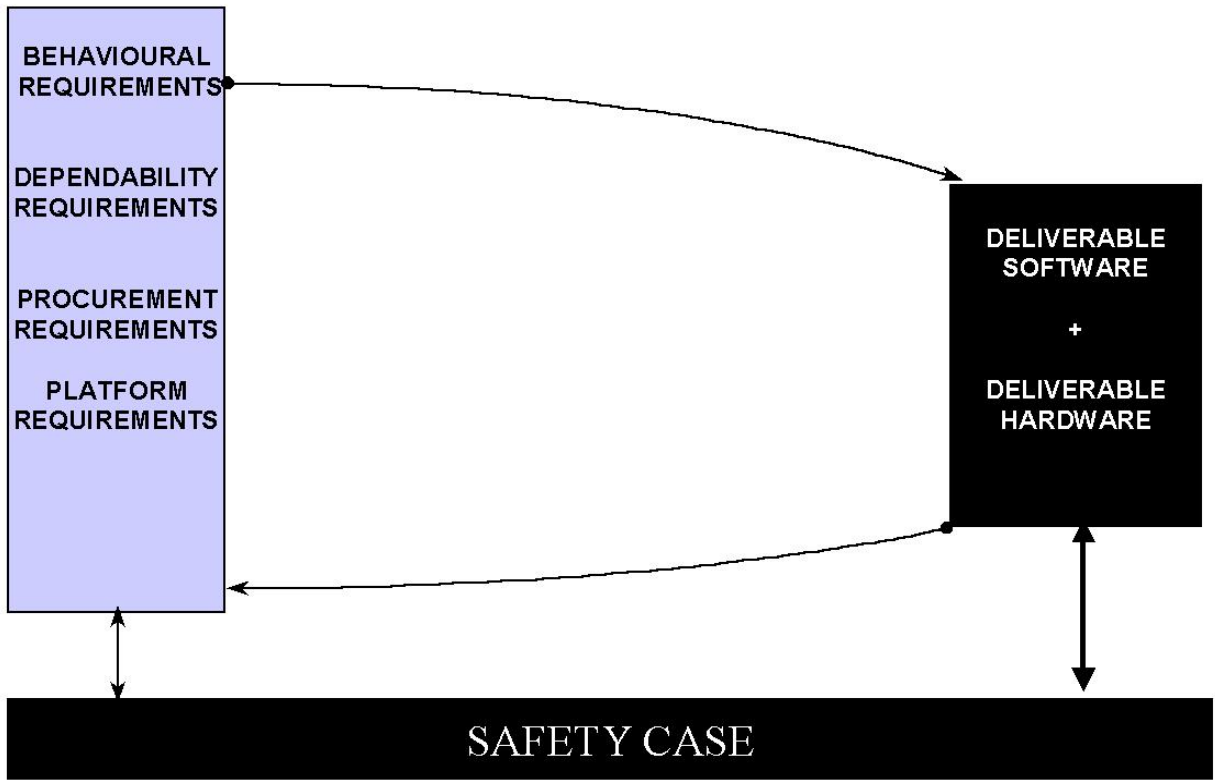

**Figure 1**: Development Process (1).

## 2. DEVELOPMENT PROCESS

In this section, we consider the development process for avionic systems. Progress is made from traditional processes through to a potential process for integrated distributed avionic systems.

*Idealistic Traditional Systems Development Process*

The traditional process of developing and procuring systems can be described, somewhat idealistically, as a collection of requirements, together with the validation of the completed system against those requirements, as

shown in Figure 1. These requirements can be broken into:
- Behavioural Requirements;
- Dependability Requirements;
- Procurement Requirements;
- Platform Requirements.

Behavioural requirements include all the expected behaviours and the prohibited behaviours. These can be represented in a variety of notations. The behavioural requirements could include mathematical models, prototypes and simulations in addition to an agreed text between the supplier and customer describing a joint expectation about the high level abstract behaviour of the system.

Dependability requirements represent a set of constraints, which must be achieved concurrently with the behavioural requirements. These can be numerical reliability targets based on historical empirical evidence from similar systems or specific safety functionality such as interlocks and fire suppression systems. In the case of the fire suppression system it can be seen that the dependability requirement has a functional and a temporal component. Dependability requirements evolve in parallel with the system refinement from high level abstract requirements to detailed implementation. At each stage there is a systematic search for new or previously undiscovered dependability requirements.

Procurement requirements constrain the development in terms of compatibility with existing fleet or making use of existing infra-structure (runways, radios, air traffic control, etc.). Increasingly, procurement policy has a significant influence on the shape and performance of the end system (e.g. "commercial-off-the-shelf" procurement policy, competitive tendering process, and international collaboration.). Lead times on the development and production of computing hardware into flight clearable LRI (line replaceable item, e.g. hardware component) and the push to reduce the overall system development time-scale has lead to more parallelism between the constituent engineering tasks.

Platform requirements are influential, on the design, as they are defined early in the lifecycle and are difficult, and potentially expensive, to change. Tolerances on weight, volume, vibration, power consumption, cooling, etc. are all drivers to the system, however their criticality is dependent on the platform. Whilst these constraints are common across the aerospace sector, the variation in criticality between a missile and a commercial aircraft leads to significantly different solutions in the different sectors of the aerospace market.

The end product of all these (potentially contradictory) requirements is an engineering compromise - a collection of computing elements consisting of dedicated hardware and software components, interacting with sensors and actuators via a variety of communication mechanisms. Verifying that this end product and the high level specification are consistent and complete is not possible by any currently known method of direct comparison. In the absence of any direct method of comparison it is necessary to consider all the intermediate engineering processes. Overall compliance is inferred from the summation of a number of stepwise compliance activities performed concurrently with the refinement of requirements into detailed implementation.

Note that *all* phases of the process are expected to contribute to the verification activity. This is illustrated in Figure 1 where all activities contribute to the Safety Case [6], which represents the collection of all evidence used in the certification of the system.

*Realistic Traditional Systems Development Process*

A realistic model of the Traditional Systems Engineering Process is shown in Figure 2. In this diagram the main intermediate engineering processes are shown. The diagram also represents the degree of parallelism, which exists between these intermediate process elements. The diagram is still idealistic to the extent that it does not show the iterations performed in order to incorporate modifications. The intermediate engineering processes illustrated in Figure 2 (in addition to those in Figure 1) are:
- Functional Requirements;
- Temporal Requirements;
- Architectural Specification;
- Equipment Specification;
- Hardware Compliance Assessment.

Functional requirements and temporal requirements are represented in a variety of notations. As the behavioural requirements are partitioned into the major elements of the system and subsequently into subsystems (which also addresses the dependability requirements), the refinement process derives timing requirements across the subsystem interfaces. The platform requirements and procurement requirements have little (if any) influence on the functional and temporal requirements as these are seen to be implementation detail.

The hardware architecture and equipment specification process is heavily driven by procurement and platform requirements. Unfortunately, the behavioural and dependability requirements are refined in parallel with the architecture and equipment specification. As a result the partitioning of hardware and software functions does not necessarily drive the hardware specification, rather the hardware which has been specified imposes a constraint on which functions can be implemented in hardware.

Delaying, the hardware architecture and equipment specification, until after the partitioning of requirements into hardware and software functions, would seem more logical. Unfortunately, the lead time from specification to equipment delivery is so long that making this a serial rather than a parallel activity would add significantly to the overall development time-scale. The implications of this are significant in terms of making the development more expensive and more obsolete by the time it enters service. In the current climate, which is trying to further shorten the overall development time-scale, removing this parallelism is not a credible option.

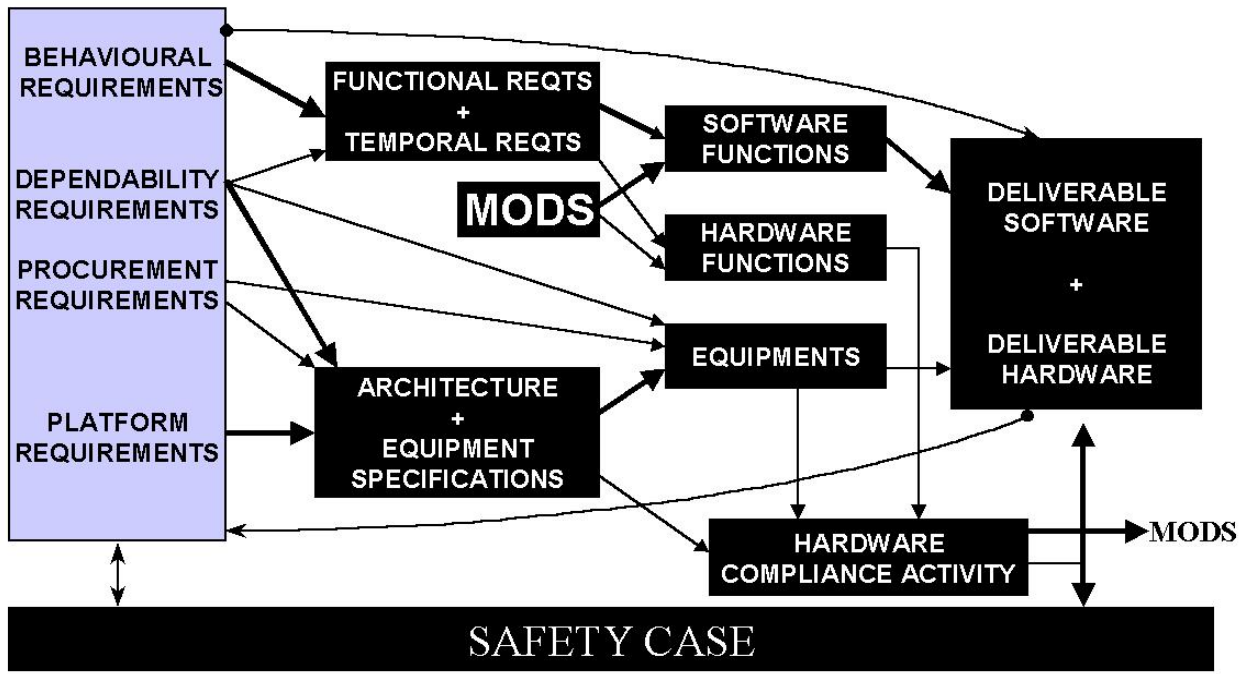

**Figure 2**: Development Process (2).

The process of partitioning from functional and temporal requirements to hardware functions and software functions is not theoretically well understood. The pragmatic approach examines the capabilities of the defined hardware and its supporting architecture and allocates functions to hardware that can support them – e.g. complex wave integration. Whatever functionality is left is allocated to software. This process of "shoe-horning" functions onto a defined standard of architecture and hardware is often felt to be less than optimal by the systems engineers and forms a source of subsequent modifications.

Components are delivered from suppliers with initially outline claims of compliance with their specification. Progressively more evidence is supplied as qualification testing is performed at significant cost in terms of money and time (burn in tests, vibration tests etc). Requirements to change hardware become progressively more expensive as the delivery process nears completion as some or all of the qualification testing may need to be repeated. Hence there is a great incentive to try and live with any limitations which may be identified.

A further problem is the integration of these components into the architecture. It is only at this time that interface inconsistencies, protocol functionality and real-time response can be assessed. Modifications to address these problems are expensive as they occur late in the development lifecycle. Once again there is pressure to try and live with any identified problems.

It is clear that given this potential for defect introduction in this process, it is difficult to produce federated architectures that ultimately reflect the expectations of the behavioural, dependability, procurement, and platform requirements.

The hardware compliance assessment activity
1. Compares individual delivered components with their specification.
2. Compares the expected behaviour of the integrated set of components against the expected performance of the overall architecture.

Expecting complete compliance on both these levels is unrealistic. Deficiencies either translate into modifications or constraints and limitations on the use/operation of the system. Note there is increasing customer resistance to these limitations and constraints.

Where modification is required the preference is to produce software modifications rather than hardware modifications. There are two main reasons for this. Firstly, software modifications are generally cheaper and have faster completion times. Secondly hardware modifications feed directly to recurring cost rather than "one off" development cost.

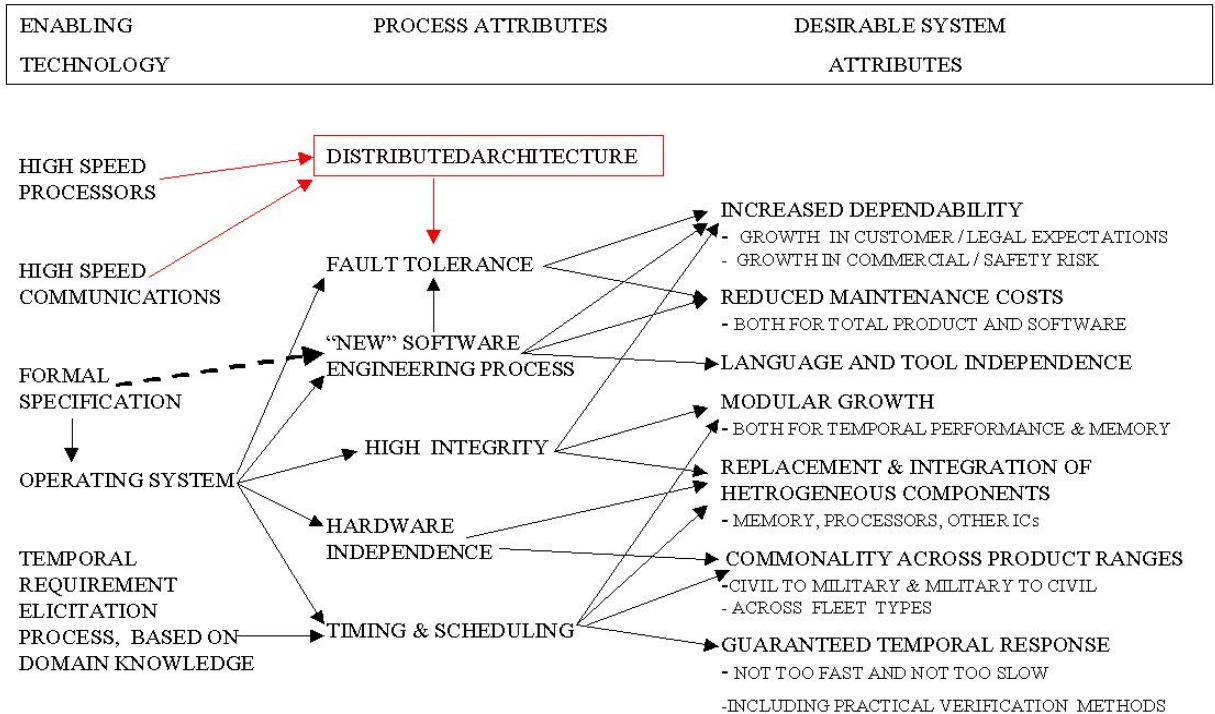

**Figure 3**: Enabling Technology.

However, the visible consequence of this is a large number of "uncontrolled" software modifications at the implementation stage. In practice, we observe that most of these modifications are directly due to the process used to specify the hardware architecture during procurement. The imposition of modifications onto the software causes the hardware specification to look artificially stable and more "right first time", rather than being recognised as a fundamental contributor to the software modification requirement.

Given that the process of defining and procuring hardware is only going to become more parallel rather than less, the issue to be considered is how do we reduce the number, severity and consequential damage of architecturally induced problems.

*Federated Architecture and the Abstract Systems Development Process – The Problems*

There are a number of problems with federated system architectures with respect to the development process:

*1. Obsolescence Management* - Staying with the hardware and architecture defined by the original development for the duration of the aerospace product lifecycle (typically 30 years) causes problems in obtaining commercial computing spares which become obsolete (typically less than 5 years). In addition to the problems of spares, an aerospace product, which is failing to exploit the latest commercially available technology, will be seen as progressively less capable when compared to newer systems.

*2. Re-qualification Costs* - The systems developed by the traditional process are qualified and certified as a single entity. This entity provides functionality from a complex interaction of interdependent software and hardware functions. Upgrading specific hardware or the complete architecture necessitates the complete re-qualification of the associated software functions even though these may not have changed.

*3. Modifications* - The hardware specification and procurement process generates a significant number of modifications. These impose a cost and time penalty. In addition it is generally recognised that modified software is more likely to contain defects than original code. This fine tuning of application software to cater for specific limitations of a particular hardware configuration is a major barrier to re-use and the adoption of software engineering practices which encourage and promote re-use.

*4. Dedicated Redundancy* - The federated architecture requires the allocation of specific functions to a particular combination of hardware and software. Dependability requirements may require specific redundancy to be provided. Depending on the criticality 2, 3 or even 4 sets of identical hardware and software must be carried as loss of a specific piece of hardware means loss of a particular function. This is inefficient, expensive and impacts directly on our ability to achieve the platform requirements

*5. Elicitation and Verification of Temporal Requirements* - There is no generally accepted process for the elicitation of timing requirements from domain specialist engineers. Whilst some requirements are expressed they are often found to be incomplete or over-specified. Over-specification can at best cause inefficient use of the available computing resource and at worst may trigger hardware upgrades late in the lifecycle, when the total available resource is found to be insufficient. Even if there is enough computing resource overall, the software may need to be re-distributed to achieve a more even spread of the processing load at a late stage of the development process. Incomplete requirements are often revealed at a late stage of the development, addressing these by modification is both costly and time consuming. Verification of timing requirements is made more difficult by the complex interdependency between the hardware and software functions and as a result almost any change to the hardware or the software requires a complete re-verification of all the timing requirements.

*Enabling Technology Constraints*

Figure 3 shows the context in which the distributed architecture sits and how it contributes to the mitigation of the problems identified in the previous section. The question could be asked that if "distribution" is such a great idea, why has it not been integrated before now into the systems development process?

As stated in section 1, there are a number of enabling technologies which are required to make a practical embedded distributed architecture viable in the aerospace sector. These include adequate hardware bandwidth, suitable fault-tolerant hardware architectures, safety analysis, timing analysis and a suitable operating system. One role of the operating system is to implement the fault tolerance requirements including reconfiguration in response to hardware failure. In addition, it must support a variety of scheduling mechanisms in a way that allows reasoned arguments to be constructed about the real-time performance of the integrated system.

The provision of such an operating system, with the required performance and evidence of high integrity, has an impact upon the software engineering process. Essentially, the process must produce applications that are independent of the underlying hardware and independent of the language/compiler/code generators used in development. Also, an elicitation process for capturing the real timing requirements from the engineers with specialist domain knowledge is required. The verification and demonstration of the achievement of the temporal requirements will make use of existing theory, but will require the development of tools to support the automation of the performance assessment. Note that the process illustrated in this section is amenable to these requirements.

## 3. ENHANCED FAULT-TOLERANT ARCHITECTURES

Fault-tolerant architectures have been used in safety-critical aerospace systems for a number of years. Traditionally, they have been based around redundant computing lanes performing identical functions, with simple (and effective) fault-detection. In this section, we examine both traditional architectures, together with architectures more amenable to a distributed integrated avionics system.

The architectures are examined from the following perspectives:
- **fault coverage** – the faults that can be detected by the architecture.
- **fault detection** – relates to how faults in the system are detected.
- **shutdown** – the hardware and software that is shutdown on a fault.
- **recovery** – the strategy used to recover from a fault.

*Traditional Federated Quadruplex / Triplex Architectures*

Figure 4 shows the typical configuration of applications in a federated avionic system. Essentially, each application has a dedicated set of sensors, processors and actuators. The figure shows one quadruplex application. Here, identical computing lanes cross-monitor the inputs to and outputs from, all lanes. Lanes falling outside some acceptable tolerance are shut down by the other lanes. This architecture has been used on military aircraft. Alternatively, a triplex architecture could be used. This architecture is used in civil aircraft, e.g. the Boeing 777 [7].

*Fault Coverage* - Quadruplex architectures offer protection from single Byzantine failures, since there are sufficient correctly functioning lanes to identify the failing lane. We note that triplex architectures cannot, in general, detect a Byzantine failure. Quadruplex, hence Byzantine, fault coverage is provided for function, data transfer and peripheral devices (e.g. sensors, actuators, power supplies etc.). Zonal coverage is also quadruplex for quadruplex systems and triplex for triplex systems.

*Fault Detection* – Fault detection is via cross-monitoring between lanes. This is commonly implemented in software – using a weighted mean system of comparing lane input/output values. Errors in sensor inputs or lane outputs are effectively masked using this system. Also, fault detection occurs by periodic "Built-In-Test" (BIT) functions, which, in general, perform basic local health monitoring functions within a lane. We note that the cross-monitoring software is (conventionally) stateless, therefore relatively error-free.

*Shutdown* - After a first non-transient fault has been detected, a quadruplex system becomes triplex by shutting down the faulty component. Thus, if the fault occurred in a sensor, that sensor input is marked as failed – the inputs from the failed sensor are subsequently triplex. This still enables voting and hence agreement to take place on input and output values. We note that Byzantine protection has been lost for the faulty component.

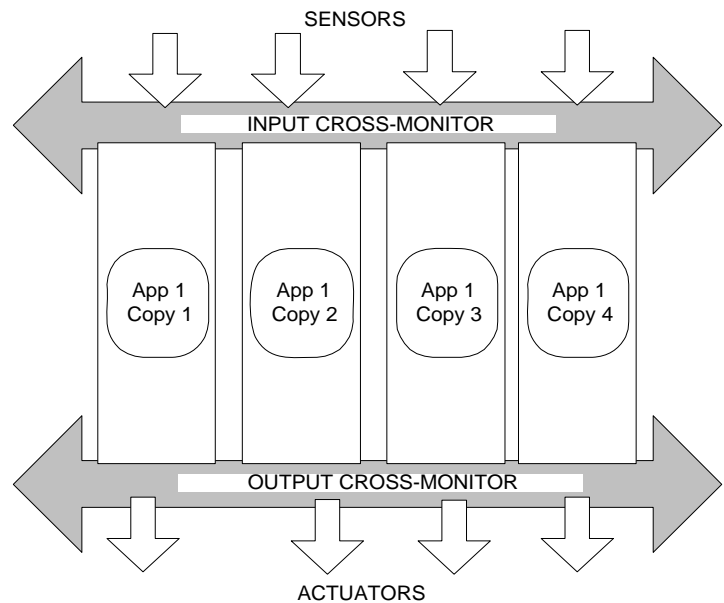

**Figure 4:** Quadruplex Architecture..

An entire lane maybe shutdown due to internal fault, e.g. failing BIT, resulting in a triplex system. Similarly, if in a triplex state when a lane fault occurs, the system reverts to duplex. Note that if a Byzantine fault had occurred the faulty lane may not have been shutdown. Hence, arguments need to be constructed to show that no credible Byzantine fault can occur. Subsequent faults are difficult to detect (within a duplex system) – if there is disagreement on input and/or output value the fault cannot be out-voted. In this situation, additional internal (within a lane) health monitoring may help.

*Recovery* – Typical quadruplex systems do not attempt any recovery via re-configuration. Instead, the failed component is given time to re-stabilise before being let back into the system. Once a previously failed component is deemed to be functioning within tolerance again, it maybe re-admitted fully into the system. This is determined by the lane comparing its inputs/outputs with those of the other lanes during the cross-monitoring process. Often, a failed component is not brought back into the system unless the pilot allows it – this prevents a transient condition occurring in vital control systems if the plane is performing some complex maneuver.

*Remarks* – The quadruplex / triplex architectures described above are suited to the static federated systems within which they are found. Since a single application has dedicated computing resource (i.e. has four lanes just performing the single application), then the consequences of lane shutdown are localised within a single application. This enables simple fault detection mechanisms to be used, based upon data cross-monitoring and voting.

*Restricted Integrated Fault-Tolerant Architecture*

Convention dictates that triplex fault-resilience is the minimum for safety-critical systems (with some noted exceptions, e.g. most digital aeroengine controllers have a duplex, or two-laned architecture). This may remove the need to use quadruplex systems, which are correspondingly more expensive to build and maintain than triplex systems. However, engineers are relatively keen to maintain adequate fault coverage after a first failure, hence the motivation for quadruplex system architectures, particularly in complex safety-critical systems such as flight control.

We noted in section 1 that the trend in system architectures is away from federated and towards distributed integrated. In this section, we consider a

triplex architecture for a limited form of integrated system:
- a computing lane contains a number of applications (c.f. traditional systems where a lane contains a *single* application).
- an application is allocated to a *single* processing resource within a lane.
- within each lane is a spare computing resource, which has (initially) no function allocated.
- after a fault, spare computing resources in the system are reconfigured to provide the functionality that has been lost.

Figure 5(a) shows an architecture that enables a limited form of re-configuration. Essentially, three lanes of identical computing elements are provided. Data-links between lanes are provided to enable cross-monitoring. On each lane, the same applications execute, with applications potentially split over one or more processing elements in a lane. We note that the allocation of application parts to processing elements is identical in each lane, at least initially. Hence triplex fault-resilience is achieved for each application. One processor on each lane is reserved as a spare.

*Fault-Coverage* – Inherently, a triplex architecture does not provide coverage against Byzantine faults. After a fault and subsequent reconfiguration, the fault coverage will remain only if all three lanes remain operational. If a lane has failed completely (e.g. power supply failure), then even if the remaining lanes can support redundant copies of the failed applications (to maintain triplex function coverage), there will only be duplex zonal fault coverage.

*Fault Detection* – Similar to traditional quadruplex, i.e. via lane cross-monitoring and "Built-In Test". However, since applications are allocated to single computing resources within a lane, fault-detection is at two granularities:
1. Lane – i.e. number of applications – if a fundamental part of the lane hardware fails, e.g. power supply.
2. Processor – i.e. single application – if the application fails (detected by the cross-monitoring) or the Built-In Test detects a local (within a computing resource) fault.

*Shutdown* – When a fault is detected, the faulty component is shut down. To reflect the granularity of fault detection, shutdown is of either a lane or processor. This enables recovery to as near a fully operational system as possible, with close to original fault-coverage with a processor fault; or may have significant reduced coverage with a lane fault. Subsequent fault effects depend upon the state that the system is in when they occur. A potential scenario is shown in Figure 5 for a triplex architecture with four processing elements and three applications (hence one spare processor in each lane). In 5(a) the initial configuration is shown; 5(b) shows a single application shutdown, 5(c) shows a lane shutdown.

*Recovery* – Recovery of failed sensors occurs in a similar manner to quadruplex systems, where they are monitored after a fault and brought back into the system after they have started to function correctly again.

Recovery at an application level occurs by re-configuration. In general, if a processor fault occurs, recovery is by configuring a spare processing resource *within the same lane* to be a "new" copy of the failed application. Clearly, depending upon the number of applications and spare processors within a lane, this could cope with a number of faults. When spare resources within a lane have been exhausted, recovery is by using spare resources within other lanes. If no resource is available, the failed application becomes duplex. Figure 5(d) shows the result of reconfiguring the failed application of 5(b) within the same lane.

For a lane failure, recovery is by utilising spare resources within other lanes to place "new" copies of all failed applications. Figure 5(e) shows the result of reconfiguring all applications within the failed lane illustrated in 5(c) – note that only two of the three applications have triplex functional fault coverage, with no application having triplex zonal fault coverage. This is potentially extendable to downgraded operation modes not covered in this paper.

*Remarks* – The restriction that an application is solely associated with a specific processor within a lane enables the simplistic fault detection mechanisms of the quadruplex architecture to be used. The major difference is that only the faulty part of a lane is shut down on fault detection. However, within this architecture, more than one application copy may be shut down by a fault – i.e. if a lane is shut down.

*Integrated Fault-Tolerant Architecture*

Taking the integrated architecture of the previous section one stage further, we no longer restrict the partitioning and allocation of applications:
- a computing lane contains a number of applications (c.f. traditional systems where a lane contains a *single* application).
- an application is composed of a number of application tasks and shared data areas.
- constituent parts of the application (i.e. tasks and shared data areas) can be allocated across multiple computing resources within the same lane.
- after a fault, spare resources are used to provide the functionality that has been lost.

The fundamental difference between this architecture and that in the previous section is that each computing resource may contain (parts of) many applications. We assume that
the *same* allocation of tasks and data is used for each lane (initially).

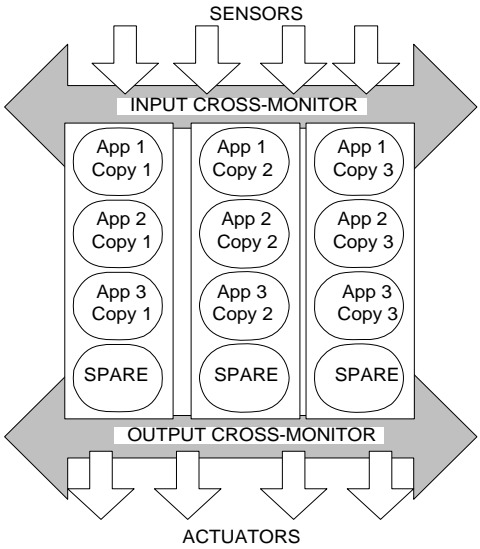
**Figure 5(a):** Initial Configuration.

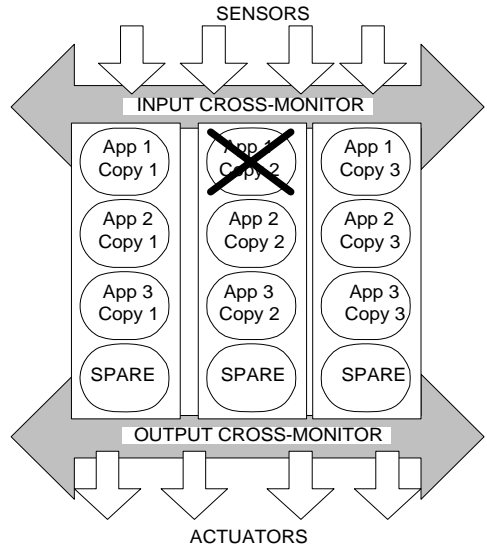
**Figure 5(b):** Single Processor Shutdown.

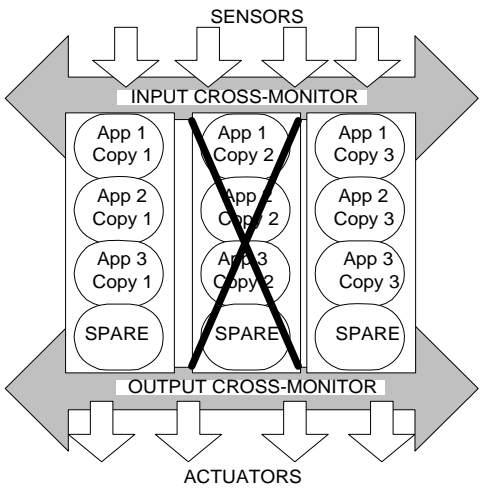
**Figure 5(c):** Single Lane Shutdown.

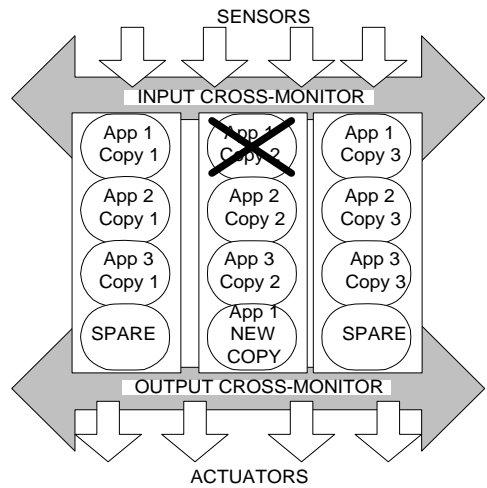
**Figure 5(d):** New Copy of Application 1.

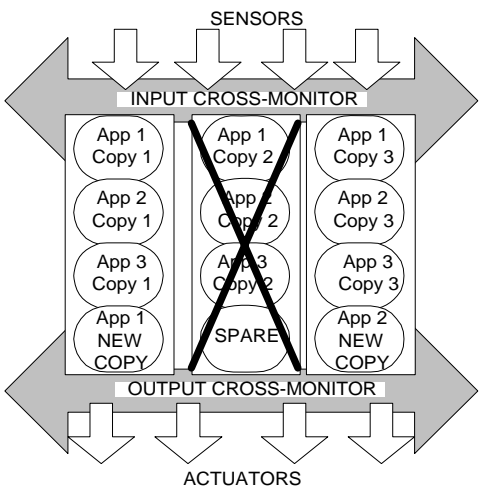
**Figure 5(e):** New (Partial) Copy of Lane.

*Fault Coverage* – This is similar to the restricted integrated architecture, whereby full triplex coverage is only maintained if all lanes are at least partially operational.

*Fault Detection* - As traditional quadruplex and the restricted integrated architecture, i.e. via lane cross-monitoring and "Built-In Test". However, now that applications are allocated to many computing resources within a lane, fault-detection is at three granularities:
1. Lane – i.e. number of applications.
2. Processor – i.e. single application.
3. Task – i.e. a task within the application.

Whilst the third form can still be provided by cross-monitoring techniques, there is a potential for the introduction of software fault-detection techniques, e.g. see [8]. Such techniques are beyond the scope of this paper.

*Shutdown* – Similarly to the restricted architecture, fault detection leads to a shutdown of the faulty component. In this architecture however, the prospect of merely shutting down a software component is raised, to correspond with the detection of a fault within a task.

*Recovery* – Recovery is via reconfiguration at a task level. A fault causes a number of effects: for a sensor fault, behaviour is as the previous architectures; other faults cause application tasks to be shutdown, from one for a task fault, through to all tasks for a lane fault. Therefore, reconfiguration requires a "new" copy of all shutdown tasks to be placed upon spare computing resources. For a task or processor fault, recovery is by configuring a spare processing resource *within the same lane* to be a "new" copy of the failed task(s). Clearly, depending upon the number of applications and spare processors within a lane, this could cope with a number of faults. When spare resources within a lane have been exhausted, recovery is by using spare resources within other lanes. If no resource is available, the tasks concerned become duplex. Note that even if one task is duplex within an application, with all other parts of the application triplex, the application only has duplex coverage.

We note that reconfiguration at a task level requires knowledge regarding load and resource requirements of the tasks – this is addressed in sections 4 and 5.

*Comparison*

In this section, we compare the traditional quadruplex, restricted integrated and integrated architectures.

Flexibility is clearly increased with the integrated architectures when compared with static federated quadruplex. However, additional run-time complexity is evident. The trade-off between the federated and integrated architectures is also dependent upon associated costs, such as potentially reduced numbers of hardware components in an integrated system and certification.

The behaviour of the architectures when a fault occurs have some similarities. The behaviour of all architectures is non- deterministic when faults occur. For example, many separate sensor failures could occur, reducing coverage on those sensors, without altering the functional fault coverage of the system. The main difference lies in the behaviour of the architectures when complete applications. Unless a complete triplex lane is shut down, the fault coverage of the re-configuring architectures is far more flexible. For example, after this single failure, the quadruplex system is reduced to triplex, whilst the re-configuring triplex architectures remain with the same coverage. A second lane shutdown for a quadruplex system will mean that system will become duplex. Comparatively, two faults in the same application under an integrated architecture results in a system with functional fault-coverage remaining triplex (although duplex zonal coverage).

Hence, there is a trade-off between the extreme fault-coverage of quadruplex systems with the more dynamic fault coverage of the distributed integrated architecture.

## 4. SOFTWARE ARCHITECTURE

In this section, we consider issues pertinent to the software architecture with respect to the restricted and fully integrated architectures introduced in section 3.

*Computational Model*

Control applications are typically structured in three phases:
1. Input – any data that is needed from sources external to the application is read, e.g. sensors, shared data etc.
2. Compute – the actual computation is performed.
3. Output – any data that needs to be passed to a destination outside the application is written, e.g. to an actuator or communications link.

We note that this form of control application is supported by the aerospace world, in both commercial forms, e.g. commercial Integrated Modular Avionics [9,10], and the military integrated avionics effort [2].

For the restricted integrated architecture, we assume that an application consists of a single cyclic executive containing a number of such tasks. For the integrated architecture, an application is composed of a number of communicating tasks. Each task has the above structure, with one task's output forming the input for the next task.

Associated with each task in an application is state data, which is persistent across successive executions of an application

*Re-Configuration of Applications*

In the distributed integrated architecture, detailed in section 3, there are a number of identical active copies of a task. When a failure occurs, there are a number of

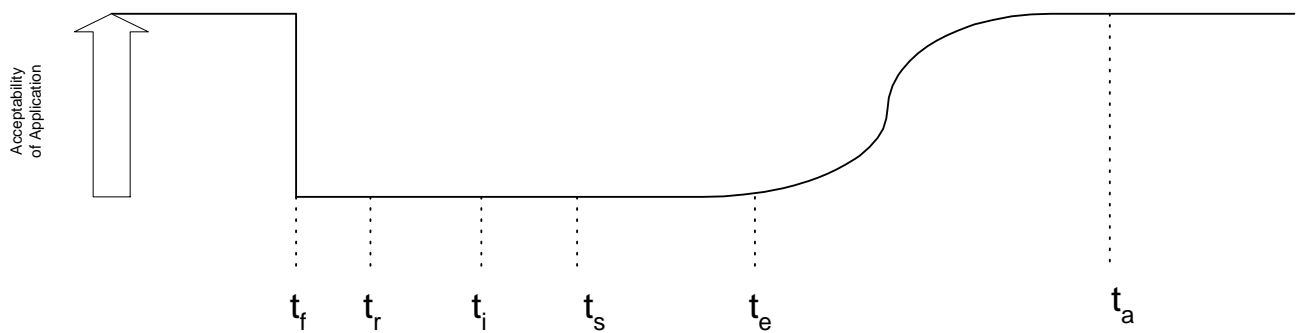

**Figure 6:** Response to Failure

phases that must be carried out to re-configure the system with "new" copies of the failed task(s):
1. Fault-detection.
2. Shutdown – this maybe for a single task (or application if a task is a complete application) or complete lane.
3. Selection – here the kernel determines the spare processor onto which the "new" copy could be placed.
4. Installation – a "new" copy of software is placed upon appropriate processor(s).
5. State Transfer – if required, state is transferred to the "new" copy.
6. Policed Execution – the "new" copy executes, although it is policed by the other remaining copies to build up a level of trust.
7. Re-admittance – the "new" copy has been accepted by the other remaining copies, and is now allowed to execute and output results normally.

This is illustrated by considering a failure, as shown in Figure 6. Here, the failure is illustrated by the acceptability of its output (i.e. state) falling when the fault is detected, and recovering after reconfiguration. The key to Figure 6:
- Time $t_f$ – failure is detected.
- Time $t_r$ - the reconfiguration commences with the selection of a suitable processing element onto which the new copy of the application is installed.
- Time $t_i$ - installation commences.
- Time $t_s$ - state transfer commences.
- Time $t_e$ - state transfer is complete, with the new application copy is eligible to be executed.
- Time $t_a$ - the new application is fully functional and re-admitted into the system.

We note that during interval $[t_f, t_a]$ the system is duplex. A subsequent fault would reduce the system to simplex. This reflects the "time-at-risk" of a secondary fault. Safety and timing analysis is required to show that this risk is acceptable.

Note that the relative distances between the time points in Figure 6 are meaningless - only the ordering is important.

One important issue is that of the state used by the new application copy. Methods of passing state to the new application are identified in the following subsections.

*State Transfer* – Some applications always need a current copy of the state, for example navigation systems often need to know where they have been before they can calculate where they are. This state may be a single "snapshot" of state, or some form of state history from which the current state maybe calculated. This state may be large.

For effective state transfer, the transferred state must be correct, else a failure may be introduced into the system. It is not sufficient to merely "vote" on the state of the other two applications in the group, as a fault cannot be masked. Also, an error may be present in the state that has yet to be detected.

*State Convergence* - Some applications are essentially stateless, or at least the state of the new copy will eventually stabilise correctly if started from afresh, merely by using new inputs. This approach is less complex than state transfer, however, the length of time from execution to re-admittance of the application $[t_e, t_a]$ may be considerably longer.

*Hybrid* - There is a trade-off between increased time at risk of second failure whilst in a duplex state (during state convergence) against the risk of commencing execution of the new application with faulty data (state transfer). Therefore, in general a hybrid solution can be used. Here, the minimum possible state is identified (offline) that is necessary for transfer. The new copy is then allowed to execute and stabilise sufficiently to pass the re-admittance test.

*Kernel*

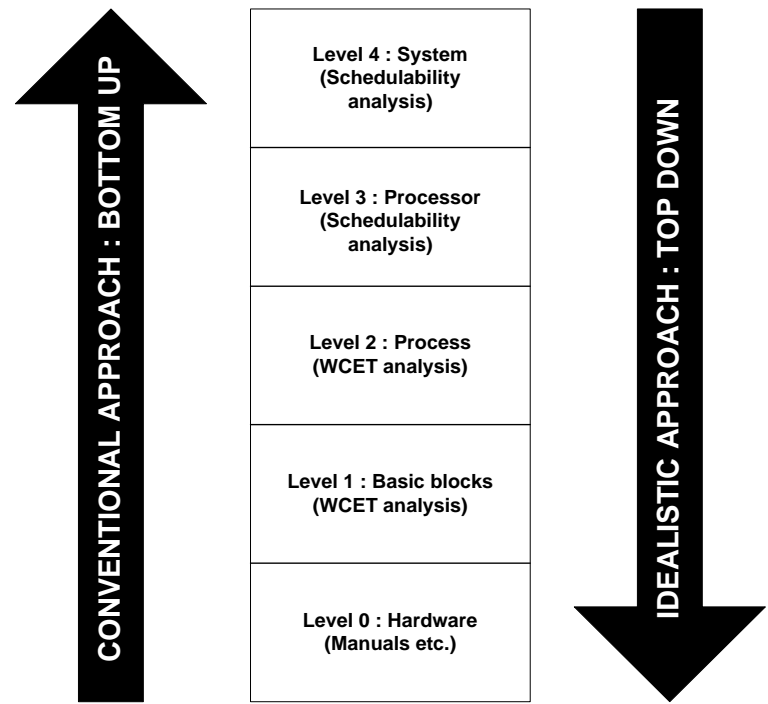

**Figure 7:** Levels of Timing Analysis

One implication of performing reconfiguration of applications is that mechanisms to support such reconfiguration are required. These are mainly functions that are performed at system initialisation time, or even offline as part of the process for writing the code to ROM:
- *Task Creation* – when a "new" task is allocated to a processor, that processor must be able to allocate memory and adjust any kernel run-time data-structures required.
- *Task Deletion* – when a single task on a processor is shutdown, it must be eliminated from the processor and any kernel level data-structures.
- *Allocation* – mechanisms for the identification of "where" to place a "new" copy of a task are required.
- *Code/Data Transfer* – mechanisms for the transfer of "new" task copies and state – normal message transfer mechanisms may be tuned for smaller data transfers than needed here.
- *Flexible Scheduling* – rather than traditional cyclic scheduling regimes, the kernel will need to support more flexible scheduling techniques that allow, more easily, the incorporation of additional tasks at run-time. Additionally, the scheduling mechanisms may allow for accelerated execution (i.e. more frequently than normal) for a task that needs to process a state history to arrive at the current state. Scheduling mechanisms are considered further in section 5.

## 5. TIMING ANALYSIS

In this section we consider the timing analysis of systems using the hardware and software architectures described in sections 3 and 4.

After an overview of timing analysis and timing requirements, we examine potential scheduling strategies for distributed systems. Then, we introduce Reservation Based Timing Analysis [5], an approach to performing offline timing analysis in a modular manner throughout the system lifecycle. The associated online scheduling techniques for distributed integrated systems are also described.

*Timing Analysis - Background*

Meeting the timing and resource usage requirements imposed upon an avionics computing system involves many aspects of the software lifecycle. Fundamentally, it is the behaviour of the system at run-time which determines whether these requirements will be met. Usually, practitioners are required to demonstrate, to customers, that timing and resource usage requirements will be met at run-time *prior* to system operation. Traditionally, this has been achieved by a combination of:
- Limited off-line analysis of timing properties of code, by *hand*.
- Exhaustive testing of the system, by exposing the system to as many combinations of input as possible, hoping to find any that lead to violation of timing requirements.

The risk with these approaches is that unless testing is complete, there may be (untested) situations that give rise to timing requirements not being met at run-time. This risk becomes greater as systems become more complex, especially when distributed systems are considered.

Alternatively, off-line timing analysis can determine whether timing requirements will be met at run-time. Rigorous mathematical analysis techniques are employed that require execution of the system software - hence can be performed without expensive system build [4]. Indeed, timing analysis can be applied at many stages of the system life-cycle.

These alternative approaches are summarised in Figure 7. Bottom-up timing analysis reflects the traditional approach, with top-down analysis the analytical approach. In both cases, five distinct layers of analysis are identified:
- **Level 0 – Hardware** - At this level, timing analysis considers the basic hardware operations that need to be quantified, e.g. the length of instructions on a processor; the physical transmission time across a communication medium. Pertinent behavioural characteristics of hardware are also noted, e.g. whether instructions on a processor are atomic, processor pipelines/caches.
- **Level 1 – Basic Blocks** - A software component can be broken into basic blocks having a single entry/exit point, with no looping, branching or calls – thus they contain no requirement of resources other than processor. At this level, timing analysis considers the processor time requirements of a basic block, e.g. the worst-case execution time is calculated by summing the worst-case execution times of component instructions.
- **Level 2 – Process** - The basic blocks identified in Level 1 may be composed into application processes by use of looping, branching and call instructions around the basic blocks. At this level, timing analysis considers the resource requirements of a process. The worst-case execution time of a process is calculated by examining the control-flow paths in the process, noting the worst-case execution times of basic blocks along the path. Also, we note other resources required by a process.
- **Level 3 – Processor** - At this level, timing analysis acknowledges the (possible) presence of multiple processes upon a single processor, which inevitably compete for resources (e.g. processor time). When multiple processes are concerned, we must take account of process scheduling within the timing analysis. Hence, the behaviour of the scheduling policy is embedded within the analysis. This is the primary reason that a timing analysis is applicable to a specific scheduling policy. Also, that scheduling policy must be used at run-time, otherwise, the timing analysis performed off-line is meaningless. One example of processor timing analysis is for priority pre-emptive scheduling, given in [4].
- **Level 4 - System** - At this level, system wide timing analysis is considered. This takes into account the true concurrency in the system, due to the multiple distinct processors. The presence of a network is considered, together with contention for network access amongst the competing processors.

As Figure 7 shows, timing analysis is layered, higher levels being dependent upon lower levels – an obvious conclusion being that all software and hardware has to be known before timing analysis can occur. However, top-down timing analysis can be performed using *estimated resource budgets* for the resources expected to be required by the software. This allows *sensitivity analysis* of the system, that is allows playing of "*what-if*" games to estimate the effects of:
- a particular software configuration;
- a particular software configuration change.

When specific resource requirements are known, they are used instead of the estimated budgets.

*Timing Requirements*

The timing requirements that may be placed upon a task or application, and can be checked by offline timing analysis, include:
- *Computational* - Each application task may have a deadline that must be met under worst-case conditions.
- *Jitter* - Tasks may have requirements on precisely when inputs are read and outputs emitted:
  - *Release Jitter* - given an event occurrence, release jitter is the difference between the earliest and latest release times of the process. The release time is when the event is recognised by the system and the appropriate process invoked, *i.e.* made runnable.
  - *Input Jitter* - at or after an event occurs, any data that is to be read by the subsequently invoked process has an interval within which it is valid. The input jitter is the difference between the time at which the data becomes valid and the latest time at which it remains valid. Input jitter is sometimes called *input validity*.
  - *Output Jitter* - when a data value is emitted by a process, output jitter represents the variation in time at which the value is emitted. Thus, output jitter is the difference between the earliest time the value can be emitted and the latest time it can be emitted. The latter is no later than the deadline of the process, by which time it must have completed execution.
- *Shared Resources* - Where resources are shared, arbitration over resource access must occur via an access protocol. The access protocol may be embedded in hardware (e.g. concurrent asynchronous memory access) or software (e.g. protocols embedded in the scheduler). However, there are two fundamental timing requirements for shared resources (both of which are usually provided by the access protocol):
- deadlock cannot occur;
- the time that a process must wait to access a resource must be bounded.
- *Mode Changes* - The system may execute in several modes, each (potentially) containing different process sets with different timing requirements. There are requirements to ensure that only the correct processes execute in a mode and that the transition between modes occurs without violating timing requirements.
- *Communications* - Where data is passed from one process to a remote process across a communications media, the time taken for the communication must be bounded.

*Scheduling Approaches for Distributed Systems*

Clearly, timing analysis is dependent upon the methods used for scheduling access to resources (i.e. processor time, communications etc.). Static, time-driven scheduling solutions have been applied in the majority of existing industrial solutions for real-time avionics and safety critical systems [17]. Whilst such solutions can provide fully deterministic behaviour, their applicability must be questioned as system complexity increases through increased levels of integration distribution. Unfavourable characteristics of such solutions are significant and include the following:
- *Inflexibility* - The final scheduling solution is not amenable to incremental change; any changes in application requirements/behaviour or resource availability imply a new static schedule.
- *Compromised application timing requirements* - True timing requirements of processes (or transactions) are compromised as all processes must fit within a common, cyclic schedule.
- *Inefficiency* – The quantity of unused resource may be significant, as when a process completes early, spare time is not reassigned to other processes. Also, when timing requirements are compromised (see above), processes are forced to run more frequently than they actually need.
- *Pessimistic worst-case estimates* - Deviation in actual run-time behaviour beyond predicted worst-case behaviour may result in information loss or cycle over-run. This encourages pessimism in design-time predictions of worst-case behaviour (computation times and rates).
- *Lack of responsiveness* - End-to-end response times may be large (compared to the total computation time involved) due to the release of successive activities based on worst-case behaviour of all predecessors. Further, any pessimism in the estimation of worst-case execution times may be directly manifest at run-time as an actual delay in end-to-end response times.
- *Lack of visibility of true precedence relationships* - In the final scheduling solution, there is no identification of true precedence relationships between processes executing on the same processor (or across multiple processors). Unless these relationships are otherwise documented at design-time, this can result in problematic modification or upgrade of the system.

Alternatively, a static fixed priority scheduling approach can be used [4]. This solves many of the shortcomings of the cyclic approaches above - for details on the trade-offs between static, time-driven scheduling and static priority-based scheduling see Audsley [12] and Locke [13].

*Reservation Based Timing Analysis*

The reservation-based approach [5] has been developed as a top down approach for timing analysis. It provides an abstract interface between application tasks and their required resources - tasks specify their time and other resource requirements in terms of a simplistic utilisation based specification. This incorporates the notion of "bounded non-determinism" - the amount of resource is stipulated, together with an interval within which the resource is required, although the precise time at which it receives that resource is determined by the scheduler. This gives the scheduler for each resource (i.e. processor, communications etc) a high degree of freedom in terms of how tasks are subsequently executed such that their timing specifications are met.

Note that given the above abstract interface between application and resource schedulers, the application is not dependent upon the specific scheduling approaches employed - it merely places a requirement upon them to provide sufficient resource within a given time frame.

Conventionally, resource reservation has been used for scheduling communications and multimedia applications [11]. These approaches have not been aimed at safety-critical applications, with little concept of off-line timing guarantees. Neither do they provide an end-to-end scheduling solution over a common set of processing and communication resources. However, resource reservation, when combined with the approaches such as fixed priority scheduling for individual processing resources, can be extended into the safety-critical domain.

We may compare the timing analyses associated with the cyclic, fixed priority and reservation based approaches:
- *Cyclic* – allows quantifiable off-line computation specification and timing guarantees; inflexible scheduling (not easily changed), hence not easily scalable.
- *Fixed Priority* - allows quantifiable off-line computation specification and timing guarantees; flexible scheduling; non-partitionable timing analysis, hence not easily scaled.
- *Reservation Based* - allows quantifiable off-line computation specification and timing guarantees;

flexible scheduling and scalability via partitionable timing analysis.

For safety-critical systems, the reservation-based approach has a number of phases [5]:
- *System Decomposition* – here the system is decomposed into constituent applications, tasks, shared data areas etc.
- *Interaction Identification* – here the interactions between various components of the system are identified, e.g. a task may need to use a specific data resource in a mutually exclusive manner.
- *Resource Reservation* - off-line timing guarantees require reservation of capacity or bandwidth on each resource required by a task. This is based upon worst-case resource requirements.
- *Offline Timing Analysis* – now, a scheduling approach is defined for each resource, with associated timing analysis used to determine whether timing requirements will be met.
- *Online Timing Analysis* – one property of the reservation approach is that all analysis is scalable and partitionable. This implies that at run-time, if a "new" component is allocated to a processor, only the timing analysis of that component need be considered. This contrasts to conventional static priority scheduling theory that requires analysis of the entire system [14].

In the remainder of this section, for brevity, we consider only the resource reservation and timing analysis phases.

*Offline Timing Analysis* - The reservation based approach assumes that a suitable offline timing analysis is available, amenable to the reservation of resources at a high level. This can be achieved using fixed priority timing analysis, assuming the use of fixed priority scheduling at run-time [4]. Also, intra-lane communications is assumed to be via a demand-driven protocol to an underlying media – i.e. as long as the total worst-case demand on the communications media is less than some maximum, then all messages have a maximum known transmission time.

For the restricted architecture, the timing analysis is simplistic. Since an application consists of a single task allocated to a dedicated processor, we only require that the computation time of the task is no more than the deadline of the task.

The fully integrated architecture is more complex in terms of timing analysis. Here we have to consider the allocation of tasks to processors within a lane, together with the effects of tasks conflicting for resources on a processor - that is the scheduling policy itself, which we have assumed to be fixed priority. We assume that priorities are assigned to tasks in a deadline monotonic fashion [15], whereby the shortest deadline task has the highest priority, the second shortest deadline has the second highest priority etc.

We use the utilisation-based test of Lui and Layland [16], noting that optimal feasibility is given in [4]. The condition to guarantee that deadlines will be met at run-time is that the total utilisation of a processor must not be greater than 69%. We note that the limitation of utilisation to 69% is not necessarily a restriction for safety-critical aerospace systems where a typical customer permits no more than 50% usage.

*Online Timing Analysis* - Inevitably, if a system reconfigures, the temporal behaviour of the system will change. However, the timing analysis approach outlined thus far caters for this.

For the restricted integrated architecture online reconfiguration causes no problem regarding timing analysis from a processor timing perspective, since an application is placed onto a "clean" processor. However, we must account for any extra communications requirement within a lane (e.g. if a "new" copy of a failed task is placed in a different lane). Thus we must maintain condition regarding total communications demand.

For the integrated architecture, during allocation of a "new" copy of a task to a processor, we must maintain the 69% utilisation condition – this will incur only a simple addition of the "new" task's utilisation to that existing on the processor. Similarly, we must check communications bandwidth in exactly the manner required for the restricted architecture.

We note that task priorities will change if a "new" task is allocated to a processor – however, this will not affect the utilisation test.

During the reconfiguration process, code and data may need to be passed across the network. This has obvious problems when other tasks within the lane are using the network for passing data as part of normal operation. We note that within the communications model assumed, the maximum available bandwidth available for transferring code/data is the difference between current load and maximum load (i.e. the difference in the previously stated communications bandwidth expression). Clearly, the available bandwidth affects the maximum time it can take to transfer the code/data.

*Online Scheduling Mechanisms* - As suggested in section 4, after a "new" copy of a task is allocated to a processor, some state may also be allocated. If this state is in a form of a history, e.g. the last 10 sets of inputs, then these must be run through the task before it has any real chance of being accepted into the system by the other "older" copies. However, to enable this, we must run the task faster than dictated by its period until it has caught up with the other copies.
One simple method for achieving this is to let the task in question use up all available spare capacity on the processor – which will be at least 31% given the requirement for a total utilisation of no more than 69% in order to facilitate offline guarantees.

## 6. Conclusions

This paper describes the observed process by which aerospace computing systems are developed. At an abstract level this process is common across all sectors of the aerospace industry (e.g. missiles, military aircraft, civil aircraft, etc.). This abstract process is unlikely to change in the foreseeable future. However, the abstract process is a major cause of hardware software integration problems and consequential lifecycle costs.

Migration to distributed architectures provides a mitigation process against the worst effects of these hardware/ software integration problems. However a number of enabling technologies need to be developed to allow the use of distributed architectures with high integrity real-time avionic applications. These include a suitable development process, distributed fault-tolerant architectures, safety-critical kernels and scalable timing analysis. This paper has examined these enabling technologies and proposed outline solutions.

One key feature of moving to distributed integrated architectures is the provision of enhanced fault-tolerance and increased availability. Traditional static approaches do not fit in well with the flexibility of the distributed integrated architecture – or at least do not utilise the full potential of the architecture. To support integrated architectures together with fault-tolerance, this paper has investigated the migration path from static fault tolerance approaches through to reconfiguration for distributed systems. In addition this paper has provided analysis to bound the temporal behaviour of fault-tolerant integrated systems in a scalable manner.

The main conclusion of this paper is that the migration to distributed avionic architectures is not only technically possible, but will have positive financial benefits to engineers adopting the approach, in terms of reduced lifecycle costs.

## 8. BIOGRAPHIES

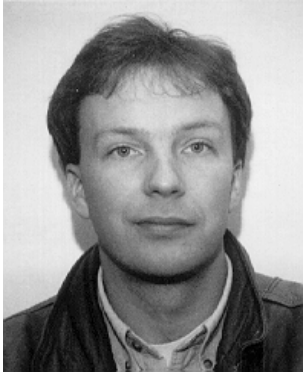

**Dr. Neil Audsley** received a BSc (Hons) and DPhil from the University of York in 1988 and 1993 respectively – the doctoral thesis considered the scheduling and timing analysis of safety-critical systems. Dr. Audsley has published numerous academic and industrial papers. He has worked with the British Aerospace Dependable Computing Systems Centre at the University of York since 1995, specialising in timing analysis and kernels for safety-critical systems.

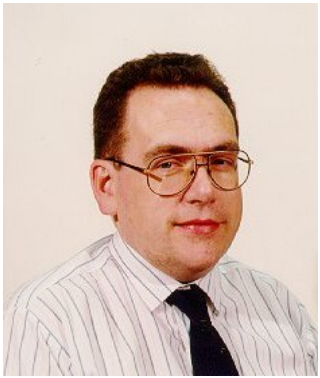

**Mike Burke** graduated from the University of Stirling in 1979. He has been a member of the British Computer Society and chartered engineer since 1992. Since joining BAe in 1979 he has worked on the specification, development and testing of embedded computing systems on a variety of Military Aircraft projects, including Tornado, Hawk and EF2000. Currently, seconded to the University of York as BAe DCSC research manager. Prior to this secondment he was the software team leader responsible for the development of the safety critical software in the EF2000 Fuel Computer.

ACKNOWLEDGEMENTS

The authors would like to thank the contributions of Steve Dawkins of the BAe Dependable Computing Systems Centre, Iain Bate of the Rolls-Royce University Technology Centre at the University of York, the BAe. Military Aircraft and Aerostructures IFPCS team, and the HISE and Real-Time research groups within the Dept. of Computer Science, University of York.